\documentclass[preprint,superscriptaddress,showpacs,prl,amsmath,amssymb,nofootinbib]{revtex4-2}
\include{setspace}
\include{dcolumn}
\usepackage{graphicx}
\usepackage{longtable}
\usepackage{multirow}
\usepackage[dvipsnames]{xcolor}
\def\bm#1{\mbox{\boldmath $#1$}}
\def\AA{nucleus-nucleus\ }
\def\dA{deuteron-nucleus\ }
\def\nA{nucleon-nucleus\ }
\begin{document}
\title{Optical model potentials for deuteron scattering off $^{24}$Mg, $^{28}$Si, $^{58}$Ni, $^{90}$Zr, $^{116}$Sn,
 and $^{208}$Pb at $\sim$100 MeV/nucleon}
\author{D. Patel} 
\affiliation{Department of Physics and Astronomy, University of Notre Dame, Notre Dame, Indiana 46556}
\author{D.~C.~Cuong} 
\affiliation{Institute for Nuclear Science and Technology, VINATOM, 179 Hoang Quoc Viet, 
 Hanoi 122772, Vietnam}
\author{K.~ B.~Howard}\thanks{Present address: Department of Science, Anne Arundel Community College, Arnold, MD 21012}
\affiliation{Department of Physics and Astronomy, University of Notre Dame, Notre Dame, Indiana 46556}
\author{U.~Garg}\email{Corresponding author; email: garg@nd.edu}
\affiliation{Department of Physics and Astronomy, University of Notre Dame, Notre Dame, Indiana 46556}
\author{Dao~T.~Khoa} 
\affiliation{Institute for Nuclear Science and Technology, VINATOM, 179 Hoang Quoc Viet, 
 Hanoi 122772, Vietnam}
\author{H.~Akimune}
\affiliation{Department of Physics, Konan University, Kobe 568-8501, Japan}
\author{G.~P.~A.~Berg}
\affiliation{Department of Physics and Astronomy, University of Notre Dame, Notre Dame, Indiana 46556}
\author{M.~Fujiwara}
\affiliation{Research Center for Nuclear Physics, Osaka University, Osaka 567-0047, Japan}
\author{M.~N.~Harakeh}
\affiliation{ESRIG, University of Groningen, 9747 AA Groningen, The Netherlands}
\author{M.~Itoh}
\affiliation{Research Center for Accelerator and Radioisotope Science, Tohoku University, Sendai 980-8578, Japan}
\author{C.~Iwamoto}\thanks{Present address: Research Center for Accelerator and Radioisotope Science, Tohoku University, Sendai 980-8578, Japan}
\affiliation{Research Center for Nuclear Physics, Osaka University, Osaka 567-0047, Japan}
\author{T.~Kawabata}
\affiliation{Department of Physics, Kyoto University, Kyoto 606-8502, Japan}
\affiliation{Department of Physics, Osaka University, Toyonaka, Osaka 560-0043, Japan}
\author{K.~Kawase}\thanks{Present address: Kansai Institute for Photon Science, National Institutes for Quantum Science and Technology, Kyoto 619-0215, Japan}
\affiliation{Research Center for Nuclear Physics, Osaka University, Osaka 567-0047, Japan}
\author{J.~T.~Matta}\thanks{Present address: Physics Division, Oak Ridge National Laboratory, 
Oak Ridge, TN 37830}
\affiliation{Department of Physics and Astronomy, University of Notre Dame, Notre Dame, Indiana 46556}
\author{T.~Murakami}
\affiliation{Department of Physics, Kyoto University, Kyoto 606-8502, Japan}
\author{M.~Yosoi}
\affiliation{Research Center for Nuclear Physics, Osaka University, Osaka 567-0047, Japan}

\date{\today}

\begin{abstract}
Angular distributions of the elastic and inelastic \dA scattering off $^{24}$Mg, $^{28}$Si, $^{58}$Ni, 
$^{90}$Zr, $^{116}$Sn, and $^{208}$Pb have been measured at a beam energy of  98 MeV/nucleon, 
with the goal of constraining the deuteron optical potential in this kinematical regime
and to extract the reduced transition probabilities for the ground-state transitions to low-lying 
excited states of these nuclei.
Two potential models were used in the analysis of the measured $(d,d)$ and $(d,d')$ data within 
the optical model and the distorted-wave Born approximation: the phenomenological optical model potential 
associated with the collective model of nuclear scattering, and the semi-microscopic double-folding 
model of the \dA potential based on a realistic density-dependent M3Y interaction. The deuteron optical 
potential and inelastic $(d,d')$ scattering form factors were calculated using these two potential models, 
allowing for a direct comparison between the potential models as well as the validation of the deduced 
$E\lambda$ transition rates.
\end{abstract}
\maketitle

\section{Introduction}
With the increasing availability of radioactive ion beams of reasonable intensities, there has been enhanced
recent interest in the investigation of giant resonances (GR), which are highly collective oscillations 
of atomic nuclei. In particular, the isoscalar giant monopole resonance (ISGMR) in nuclei far from 
the stability line has the potential to make significant strides in our understanding of nuclear
incompressibility, $K_{\infty}$, and especially the asymmetry term, $K_{\tau}$ \cite{garg-colo}. Such
measurements have to be performed in inverse kinematics, with deuterium and helium gases being the best targets 
available so far. While inelastic scattering of $\alpha$ particles has been the mainstay 
of such studies for a long time now \cite{garg-colo}, there has not been much work with deuterons 
since the mid-70's primarily because of various experimental constraints \cite{france1, france2}. 
On the other hand, most of the measurements of GR with radioactive ion beams so far have employed 
active-target time projection chambers (AT-TPC) with deuterium as the component gas 
\cite{ganil1, ganil2, ganil3, ganil4}. Because of paucity of GR data with deuterons, it was important to 
validate in known cases the results obtained in $(d,d')$ measurements via direct comparison with results 
of inelastic $\alpha$ scattering. Such a detailed investigation was carried out by measuring small-angle 
inelastic deuteron scattering off $^{116}$Sn and $^{208}$Pb, and it was established that the extracted 
ISGMR strength distributions, using a multipole decomposition analysis similar to that done 
for inelastic $\alpha$ scattering, agree very well with those deduced from the $(\alpha,\alpha')$ 
data \cite{dpatel2013}. 

Such analyses hinge upon obtaining good-quality optical model (OM) parameters from elastic 
scattering data. The OM potential is widely used to generate wave functions for elastic scattering. These wave 
functions serve, in turn, as the door waves, widely known as the distorted waves, for the analysis of other direct 
reaction channels within the distorted-wave Born approximation (DWBA) or coupled-channel formalism. 
The inelastic scattering leading to the excitation of collective states of the target nucleus is essential
for the determination of the underlying nuclear structure properties from the measured angular distributions.  
The nuclear interaction potential is, however, inherently complicated. Therefore, the phenomenological optical 
potential (OP) of a simple functional form is often employed in the OM and DWBA calculations to describe 
direct nuclear reactions and to extract nuclear properties from the angular distribution data. It represents 
a simple ``effective'' interaction used in the collective model of nuclear scattering \cite{Tamura65}
to describe both the elastic and inelastic scattering channels.

For elastic \dA scattering, the existing OM studies of the deuteron OP are mainly based on the phenomenological 
potential model using the Woods-Saxon (WS) functional form as described, for example, in Ref.~\cite{daehnick1980}. 
This simple OP was successfully used to describe the elastic as well as inelastic angular distributions of deuteron 
scattering off heavy nuclei (A $\gtrsim$ 40), based on the collective model of nuclear scattering \cite{Tamura65}. 
However, the collective model approach seems to overestimate the (one-step) transition probabilities from 
the ground state to the low-lying excited states of some light nuclei. For example, as noted in 
Ref.~\cite{korff2004}, the OM parameter set that gives the best fit of the elastic angular distribution 
requires 40\% lower value of the deformation parameter $\beta_2$, as compared to the adopted value, 
to reproduce the experimentally observed 2$^+_1$ angular distribution in the nucleus $^{16}$O. 

In this work, we report on the elastic and inelastic \dA scattering measurements at the beam energy of 98 MeV/nucleon
for several targets, ranging from light- to medium- and heavy-mass nuclei. The measured $(d,d)$ and $(d,d')$ 
scattering data have been analyzed within the OM and DWBA, respectively, using both the WS phenomenological 
potential model as well as a semi-microscopic \dA potential obtained in the double-folding model  
\cite{Sat79,Bra97,Kho00}. The extracted OM parameters are expected to find use in analyzing the giant 
resonance data with radioactive ion beams.  

The experiments were performed at the ring cyclotron facility of the Research Center for Nuclear Physics 
(RCNP), Osaka University, Japan. The $E_d=196$ MeV deuteron beam was scattered off six highly 
enriched (more than 90\%) self-supporting targets listed in Table~\ref{tab:Thickness}. Elastic and inelastic 
\dA scattering measurements were made over an angular range of $\theta_{\rm lab}\sim 3.5^\circ - 32^\circ$. 
Each experimental angular opening ($\sim 2^\circ$-wide) was subdivided into three parts for the analysis, each corresponding to a solid angle of 0.42 msr.
\begin{table}[h!]
		\begin{center}
		\caption{Target specifications}	\label{tab:Thickness}
		\begin{tabular}{|c|c|c|c|} \hline
Target & Thickness  & Target & Thickness \\
 & (mg/cm$^2$) &  & (mg/cm$^2$) \\ \hline
 $^{24}$Mg & 50.0 & $^{90}$Zr & 4.2 \\
 $^{28}$Si & 58.5 & $^{116}$Sn & 10.0\\
 $^{58}$Ni & 1.5 & $^{208}$Pb & 10.0\\ \hline
			\end{tabular}
			\end{center}
\end{table}

The scattered particles were momentum analyzed by the magnetic spectrometer Grand Raiden and focused onto 
the focal-plane detector system \cite{Itoh2003} consisting of two multi-wire drift chambers (MWDC) and 
two plastic scintillators \cite{MWDCreport99}. The time-of-flight and energy-loss techniques were used for the 
identification of the scattered particles. Grand Raiden was used in the double-focusing mode in order to identify 
and eliminate practically all instrumental background from the final spectra \cite{dpatel2013}. Particle tracks 
were reconstructed using the ray-tracing technique described in Refs.~\cite{Itoh2003, Li2010}. This, in turn, 
allowed for the reconstruction of the scattering angle. The experimental angular resolution  was 
$\sim$ 0.15$^\circ$, including the nominal broadening of the scattering angle due to the emittance 
of the beam and the multiple Coulomb scattering effects. Further details of the experimental and data analysis 
procedures have been provided in Ref. \cite{dpatel2013,dpthesis}.

\section{DWBA analysis based on the phenomenological OM potential}\label{sec2}
\begin{figure}
 \begin{centering}\vskip -0.5cm\hskip -0.5cm
 \includegraphics[width=1.0\linewidth]{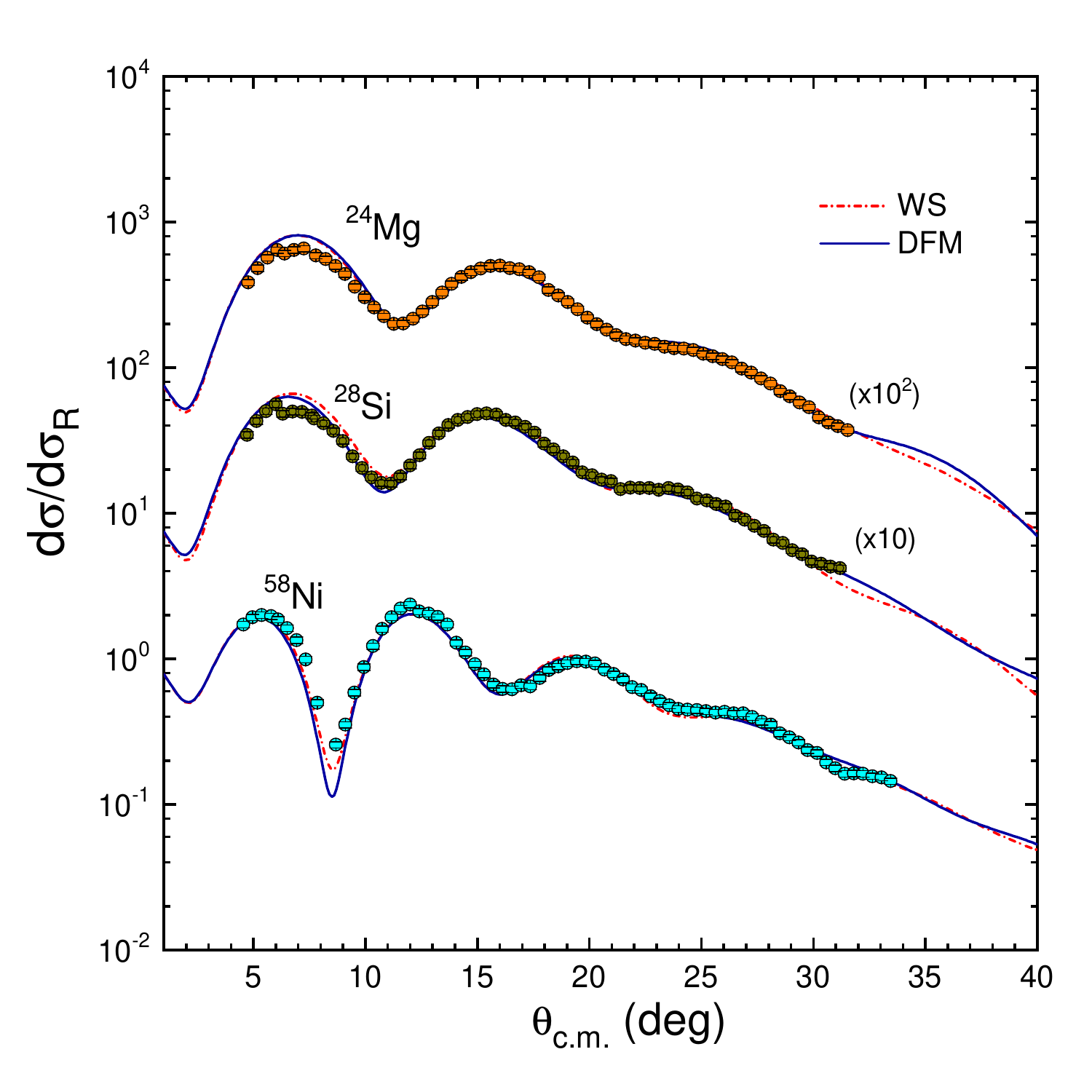}\vskip -0.5cm 
 \caption{Elastic $(d,d)$ scattering data measured at $E_d=196$ MeV (in ratio to the corresponding 
Rutherford cross sections) for $^{24}$Mg, $^{28}$Si,  and $^{58}$Ni targets  (solid points). 
The dashed and solid lines are the OM results given by the phenomenological OP 
(\ref{eq1})-(\ref{eq6}) and hybrid folded OP (\ref{df1})-(\ref{df6}), respectively.} \label{f1}
\end{centering}
 \end{figure}

\begin{figure} 
\begin{centering}\vskip -0.5cm\hskip -0.5cm
 \includegraphics[width=1.0\linewidth]{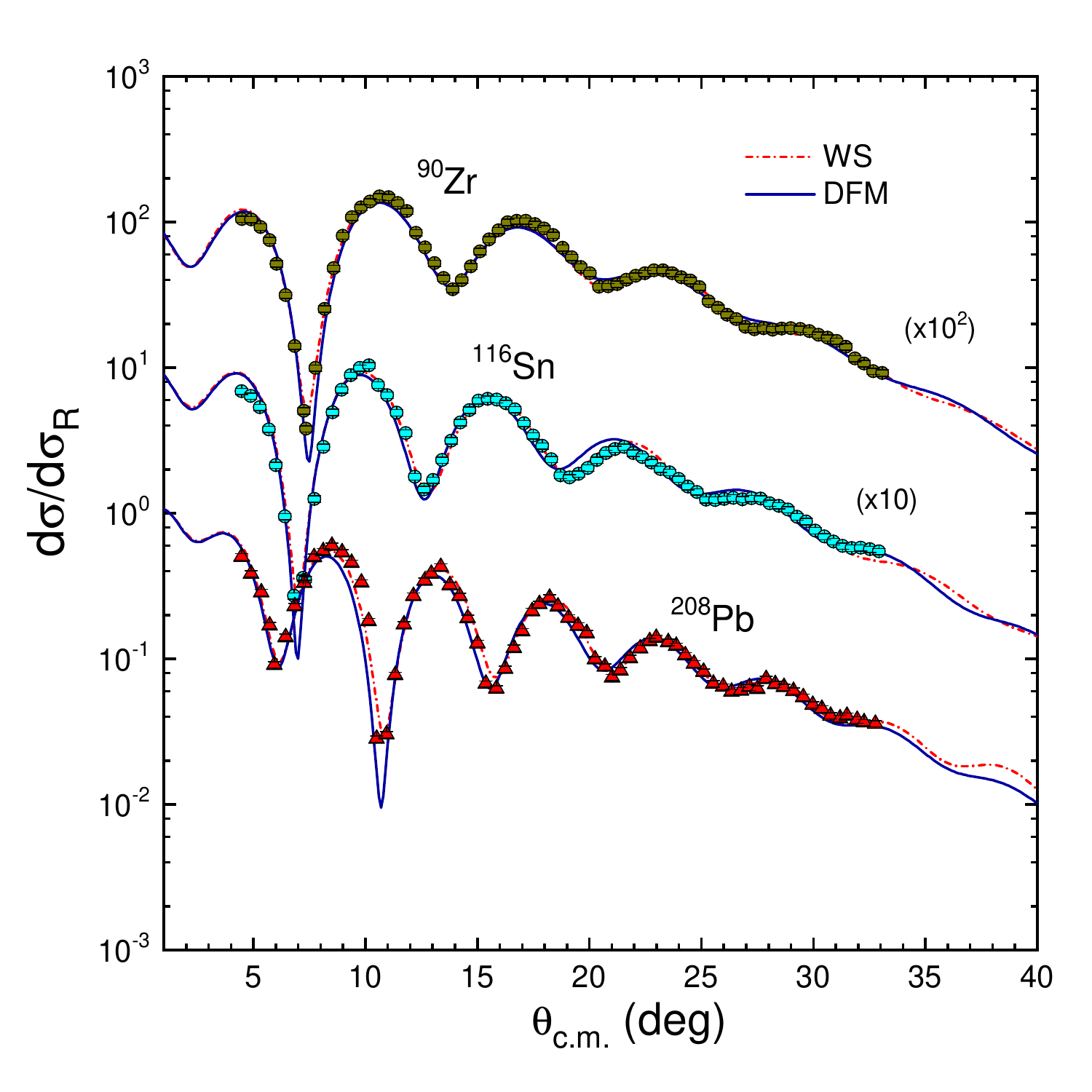} \vskip -0.5cm
 \caption{The same as Fig.~\ref{f1} but for $^{90}$Zr, $^{116}$Sn,  and $^{208}$Pb 
 targets.} \label{f2}
\end{centering}
 \end{figure}

The two available global \dA optical potentials were developed by Daehnick \textit{et al.} 
\cite{daehnick1980} and Bojowald \textit{et al.} \cite{bojo1988} in the 1980s using the 
phenomenological WS form for both the real and imaginary OP, covering the mass range 
of 27 $\leq$ A $\leq$ 238 and energy range of $E_d \sim10$ MeV--90 MeV. The deuteron 
carries one unit of spin ($\bm{s}=\bm{1}$) in its ground state, and this requires the inclusion 
of a real spin-orbit term $V_{\ell s}$ into the OP. For the imaginary OP, in addition to the volume 
absorption $W$, an imaginary surface term $W_D$ is also included to account for the surface 
absorption which is significant due to the deuteron breakup \cite{Rawitscher74,Austern87}. 
Thus, the total OP is determined explicitly as
\begin{equation}
U(r) = V(r) + iW(r)+iW_D(r)+V_{\ell s}(r)(\bm{\ell}\cdot\bm{s})+V_C(r) ,
\label{eq1}
\end{equation}
\begin{eqnarray}
\mbox{where}\ V(r)&=&-Vf(r,r_{V},a_{V}), \label{eq2} \\  
 W(r)&=&-Wf(r,r_W,a_W), \label{eq3} \\ 
 W_D(r)&=&4a_DW_D{\frac{d}{dr}}f(r,r_D,a_D), \label{eq4} \\
V_{\ell s}(r)&=&V_{\ell s}\left[\frac{\hbar}{m_{\pi}c}\right]^{2}
 \frac{1}{r}\frac{d}{dr}f(r,r_{\ell s},a_{\ell s}), \label{eq5}
\end{eqnarray}
and $V_C(r)$ is the Coulomb potential of a uniformly charged sphere of radius 
$R_C=1.3~ A^{1/3}$ fm. The functional form $\textit{f}$ is chosen in the WS form 
for all the terms
\begin{equation}
f(r,r_i,a_i)=\left[1+\exp\left(\frac{r-r_iA^{1/3}}{a_i}\right)\right]^{-1}. \label{eq6}
\end{equation}
The phenomenological OM potential given by Eqs. (\ref{eq1})-(\ref{eq6}) is used in the OM analysis 
of the measured elastic $(d,d)$ 
scattering data, and to generate the distorted waves for the DWBA description of the inelastic $(d,d')$ 
scattering data.  The associated OM parameters were obtained from a $\chi^2$-minimization 
fit to the  elastic $(d,d)$ data using the code ECIS97 \cite{ecis}, and they are 
given explicitly in Table~\ref{t2}. The global deuteron OPs \cite{daehnick1980,bojo1988} were used for 
the parameter initialization in the $\chi^2$ search. 
\begin{table*}[h]
	\begin{center}\vskip -0.2cm
	\caption{Best-fit parameters of the phenomenological OP (\ref{eq1})-(\ref{eq6}). Because of spin convention, 
	the $V_{\ell s}$ value must be divided by 2 when used in the numerical input of the code ECIS97 \cite{ecis}. 
	The errors were deduced from the weight of each parameter in the covariant multi-parameter $\chi^2$-search, 
	with $r_V$ and $V_{\ell s}$ kept fixed during the search.}
	\label{t2}\vskip 0.5cm
\begin{tabular}{|c|c|c|c|c|c|c|} \hline
Target & $^{208}$Pb & $^{116}$Sn & $^{90}$Zr & $^{58}$Ni & $^{28}$Si & $^{24}$Mg \\
\hline
$V$ (MeV) & $46.54\pm 0.01$ & $44.33\pm 0.01$ & $42.95\pm 0.13$ & $39.07\pm 0.14$ 
 & $35.58\pm 0.13$ & $31.92\pm 0.13$ \\
$r_V$ (fm) & 1.18 & 1.18 & 1.18 & 1.18 & 1.18 & 1.18 \\
$a_V$ (fm) & $0.938\pm 0.001$ & $0.911\pm 0.001$ & $0.997\pm 0.013$ & $0.914\pm 0.004$ 
 & $0.911\pm 0.013$ & $0.977\pm 0.004$ \\ \hline
$W$ (MeV) & $20.59\pm 0.01$ & $20.87\pm 0.03$ & $20.20\pm 0.20$ & $21.41\pm 0.20$ 
 & $22.67\pm 0.13$ & $24.39\pm 0.14$ \\
$r_W$ (fm) & $1.160\pm 0.001$ & $1.070\pm 0.001$ & $1.060\pm 0.013$ & $1.100\pm 0.019$ 
 & $0.850\pm 0.001$ & $1.000\pm 0.012$ \\
$a_W$ (fm) & $0.361\pm 0.002$ & $0.670\pm 0.002$ & $0.538\pm 0.012$ & $0.456\pm 0.004$ 
 & $0.420\pm 0.005$ & $0.501\pm 0.003$ \\
\hline
$W_D$ (MeV) & $7.00\pm 0.01$ & $7.00\pm 0.01$ & $7.90\pm 0.19$ & $7.60\pm 0.19$ 
 & $7.95\pm 0.04$ & $7.50\pm 0.14$ \\
$r_D$ (fm) & $1.230\pm 0.001$ & $1.110\pm 0.001$ &  $1.100\pm 0.019$ &  $1.050\pm 0.020$ 
 &  $1.000\pm 0.004$ &  $1.022\pm 0.005$ \\
$a_D$ (fm) & $0.790\pm 0.001$ & $1.080\pm 0.001$ & $0.997\pm 0.006$ & $1.030\pm 0.015$ 
 & $0.980\pm 0.001$ & $0.920\pm 0.003$ \\
\hline
$V_{\ell s}$ (MeV) & 4.22 & 4.22 & 4.22 & 4.22 & 4.22 & 4.22 \\
$r_{\ell s}$ (fm) & $1.150\pm 0.001$ & $1.130\pm 0.001$ & $1.200\pm 0.001$ & $1.190\pm 0.017$ 
 & $1.130\pm 0.015$ & $1.163\pm 0.003$ \\
$a_{\ell s}$ (fm) & $1.230\pm 0.001$ & $1.110\pm 0.013$ & $0.985\pm 0.002$ & $1.110\pm 0.012$ 
 & $1.110\pm 0.020$ & $1.164\pm 0.013$ \\
\hline
\end{tabular}
\end{center}
\end{table*}
The OM results obtained using the best-fit OP parameter set are shown as dashed lines in Figs.~\ref{f1} 
and \ref{f2}, in comparison with elastic $(d,d)$ scattering data measured at $E_d=196$ MeV for the targets 
under study.

The DWBA analysis of inelastic deuteron-nucleus scattering data is restricted to the direct one-step 
excitation of a collective state of the target, neglecting the contributions of indirect two- or three-step 
transitions and the channel coupling effects. The $(d,d')$ scattering cross section is calculated 
using the following DWBA inelastic scattering amplitude \cite{Satchler83}:
\begin{equation}
T_{\rm DWBA}=\int\left[\chi^-(\bm{k}',\bm{r}')\right]^* \langle dA' |V_{d-A}|dA\rangle 
\chi^+(\bm{k},\bm{r})d^3rd^3r', \label{eq7}
\end{equation}
where the distorted waves $\chi^\pm$ are generated by the OP (\ref{eq1})-(\ref{eq6}). 
The transition matrix element of the \dA interaction $\langle dA' |V_{d-A}|dA\rangle$ is dubbed 
as the inelastic scattering form factor (FF). A widely adopted method is to obtain the \emph{nuclear} 
inelastic scattering FF by radially deforming nuclear part of the OP, based on the collective vibrational 
(or rotational) model of nuclear scattering \cite{Tamura65}. For the $(d,d')$ scattering to a 
$2^\lambda$-pole collective excited state of  target, the inelastic scattering FF is determined as
\begin{eqnarray}
F_\lambda(r)&=&F^{(\lambda)}_N(r) +F^{(\lambda)}_{\ell s}(r)(\bm{\ell}\cdot\bm{s})
+F^{(\lambda)}_C(r) ,  \label{eq8} \\
F^{(\lambda)}_N(r)&=&\delta^{(N)}_\lambda\frac{d}{dr} [V(r) + iW(r)+iW_D(r)], \label{eq9}\\
F^{(\lambda)}_{\ell s}(r)&=&\delta^{(N)}_\lambda\frac{d}{dr}V_{\ell s}(r). 
\label{eq10} 
\end{eqnarray}    
Prescription (\ref{eq9})-(\ref{eq10}) is known as the deformed optical potential (DOP) 
method to generate the inelastic nuclear scattering FF.  The spin-orbit contribution (\ref{eq10}) 
to the total $(d,d')$ cross section is minor, but it helps to better reproduce the observed 
oscillation pattern of  inelastic $(d,d')$ cross section for medium mass targets.    
For an isoscalar excitation, the nuclear deformation length $\delta^{(N)}_\lambda$ is usually
assumed to be the same as the proton deformation length $\delta^{(p)}_\lambda$, determined  
from the reduced transition rate $B(E\lambda\uparrow)$ \cite{Satchler83} as 
\begin{equation}
 \delta^{(p)}_\lambda=\frac{4\pi\sqrt{B(E\lambda\uparrow)/e^2}}{3ZR_A^{\lambda-1}},\
 R_A=1.3A^{1/3}\ \mbox{fm}. \label{eq11}
\end{equation}
The Coulomb inelastic FF in Eq.~(\ref{eq8}) can be approximately obtained in a 
``model independent" form \cite{Satchler83}:
\begin{equation}
 F^{(\lambda)}_C(r)=\frac{4\pi\sqrt{B(E\lambda\uparrow)}\ e}
{(2\lambda+1)r^{\lambda+1}}.  \label{eq12}
\end{equation}
Thus, the OP parameters (\ref{eq1})-(\ref{eq6}) and the reduced transition rate 
$B(E\lambda\uparrow)$ of the target excitation are the main inputs for the DWBA 
analysis of inelastic $(d,d')$ scattering using the collective model of nuclear scattering 
\cite{Tamura65} associated with the phenomenological OP). 
The DWBA calculation of the $(d,d')$ scattering cross section has been done 
using two choices of the $B(E\lambda)$: 

i) $B(E\lambda\uparrow)$ is fixed to the experimentally adopted transition rates 
 \cite{Raman01,nndc,Kibedi02}, denoted as $B(E\lambda)_{\rm adopted}$  in Table~\ref{t3};

ii)  $B(E\lambda\uparrow)$ is deduced from the best DWBA fit of the calculated $(d,d')$ cross
section to the $(d,d')$ scattering data and denoted as $B(E\lambda)_{\rm DOP}$ or 
$B(E\lambda)_{\rm DFM}$ in Table~\ref{t3}.  

\begin{table}[h]
\begin{center}
\caption{Experimentally adopted values for the reduced electric transition rates 
	$B(E2\uparrow)$ \cite{Raman01,nndc}, $B(E3\uparrow)$ \cite{Kibedi02}, and the best-fit 
	transition rates $B(E\lambda)_{\rm DOP}$ and $B(E\lambda)_{\rm DFM}$ deduced from the 
	DWBA analysis of  the $(d,d')$ data using the inelastic FF based on the phenomenological	OP 
	and that based on the hybrid folded \dA potential, respectively. The errors were estimated from 
	those of the measured angular distributions. } 
\label{t3}\vskip 0.5cm
\begin{tabular}{|c|c|c|c|c|c|} \hline
Target & $E_x$ & $\lambda^\pi$ & $B(E\lambda)_{\rm adopted}$ &$B(E\lambda)_{\rm DOP}$ & 
 $B(E\lambda)_{\rm DFM}$ \\
& (MeV)& &($e^2$ b$^{\lambda}$)&($e^2$ b$^{\lambda}$) & ($e^2$ b$^{\lambda}$) \\ \hline
$^{24}$Mg & 1.368 & $2^+$& 0.0432 \textsl{(11)} &  0.0284  \textsl{(14)}& 0.0410  \textsl{(21)}  \\
$^{28}$Si & 1.779 & $2^+$& 0.0326 \textsl{(12)} & 0.0196  \textsl{(10)} & 0.0326  \textsl{(16)} \\
$^{58}$Ni &1.454 & $2^+$ & 0.0695 \textsl{(20)}& 0.0695 \textsl{(35)}& 0.0695  \textsl{(35)} \\
$^{90}$Zr & 2.186 & $2^+$ & 0.061 \textsl{(4)} &0.0641  \textsl{(32)}& 0.0641  \textsl{(32)} \\
          & 2.748 & $3^-$ & 0.098 \textsl{(5)}$^a$ & 0.0510  \textsl{(37)}& 0.0640  \textsl{(32)} \\
          &  &   & 0.037 - 0.079 $^b$ &  &  \\
$^{116}$Sn & 1.293 & $2^+$ &0.209 \textsl{(6)} &0.241  \textsl{(12)}& 0.240  \textsl{(12)} \\
          & 2.266 & $3^-$ & 0.132 \textsl{(18)} $^c$ & 0.142  \textsl{(7)} & 0.160  \textsl{(8)} \\
          &  &   & 0.112 - 0.202 $^b$ &  &  \\				
$^{208}$Pb & 2.610 & $3^-$ & 0.611 \textsl{(9)} $^c$ & 0.611  \textsl{(31)}& 0.642  \textsl{(32)} \\
          &  &   & 0.419 - 0.836 $^b$ &  &  \\  \hline
\end{tabular} \\ 
\vskip 0.5cm 
 %
 $^a$adopted from $(e,e')$ data \\
 $^b$ from inelastic nucleon- and light-ion scattering data \\
 $^c$ adopted from Coulomb excitation data \\
 \end{center}
\end{table}

 \begin{figure}
 \begin{centering}\vskip -0.5cm
 \includegraphics[width=1\linewidth]{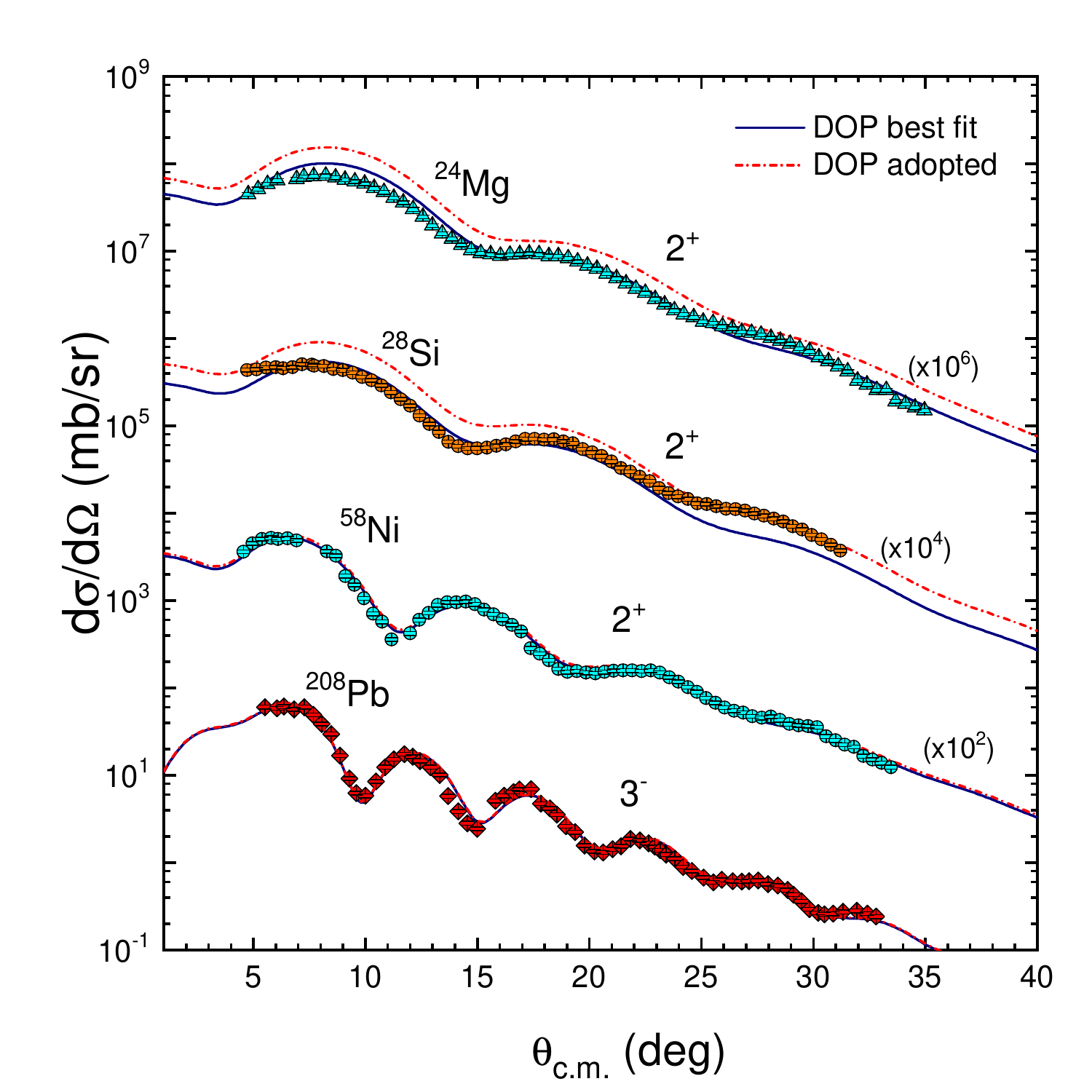}\vskip -0.2cm
 \caption{The inelastic $(d,d')$ scattering data measured at $E_d=196$ MeV (solid points) 
for the excitation of 2$^+_1$ states of $^{24}$Mg, $^{28}$Si, $^{58}$Ni targets, and 3$^-_1$ 
state of $^{208}$Pb target. The DWBA results given by the collective model FF (\ref{eq8})-(\ref{eq12}) 
based on the adopted $B(E\lambda)_{\rm adopted}$ and best-fit $B(E\lambda)_{\rm DOP}$ values 
(see Table~\ref{t3}) are shown as dashed and solid lines, respectively.} \label{f3}
 \end{centering}
 \end{figure}
The DWBA description of  inelastic $(d,d')$ scattering data measured for the excitation of 2$^+_1$ states 
of $^{24}$Mg, $^{28}$Si, $^{58}$Ni targets, and 3$^-_1$ state of $^{208}$Pb target given by the 
collective-model inelastic FF (\ref{eq8})-(\ref{eq12}) is shown in Fig.~\ref{f3}. As can be seen, 
the 3$^-_1$ angular distribution measured for $^{208}$Pb target is reproduced well with the adopted 
$B(E3)_{\rm adopted}$ value \cite{Kibedi02}. However, the DWBA description of the measured 2$^+_1$ 
cross section using the adopted $B(E2)$ value, $B(E2)_{\rm adopted}$, seems to get worse for light $^{24}$Mg and 
$^{28}$Si nuclei, where the calculated $(d,d')$ cross section overestimates data at small angles, and then becomes non-oscillatory for angles above $\sim$ 25$^{\circ}$. The best DWBA fit to inelastic $(d,d')$ data given
by the collective-model FF requires $B(E2)_{\rm DOP}$ value of around 40\% lower than $B(E2)_{\rm adopted}$ 
value. This likely indicates a deficiency of the collective model of nuclear scattering \cite{Tamura65} based 
on the phenomenological OP (\ref{eq1})-(\ref{eq6}) for light-mass nuclei, as also found earlier for $^{16}$O 
in Ref.~\cite{korff2004}.  
\begin{figure}[h]
 \begin{centering}\vskip -0.5cm
 \includegraphics[width=1\linewidth]{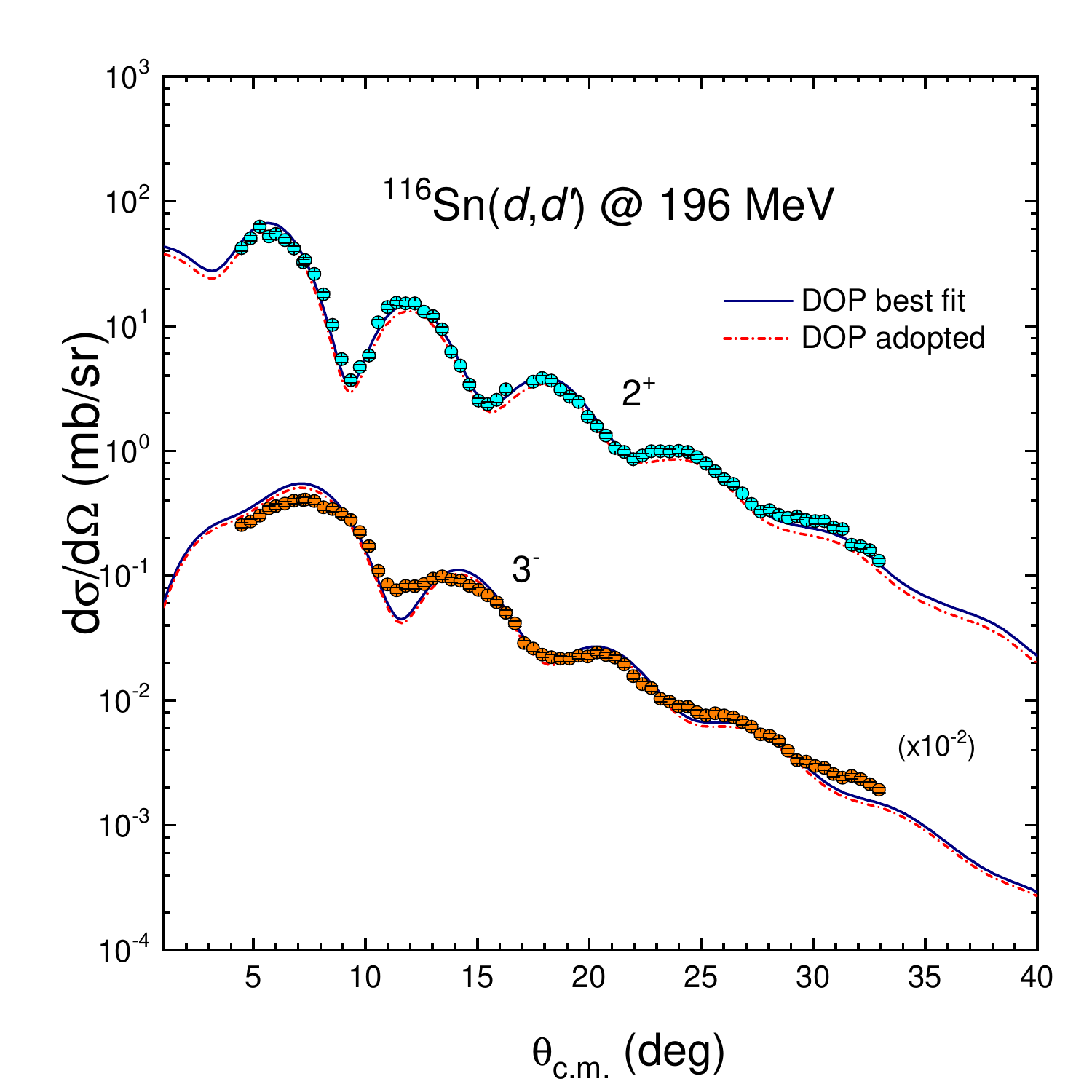}\vskip -0.2cm
 \caption{Inelastic $^{116}$Sn$(d,d')$ scattering data (solid points)  measured at $E_d=196$ MeV 
for the excitation of the 2$^+_1$ and 3$^-_1$ states of $^{116}$Sn target. The DWBA results given 
by the collective model FF (\ref{eq8})-(\ref{eq12}) based on the adopted $B(E\lambda)_{\rm adopted}$ 
and best-fit $B(E\lambda)_{\rm DOP}$ values (see Table~\ref{t3}) are shown as dashed and solid lines, 
respectively.} \label{f4}
 \end{centering}
 \end{figure}
\begin{figure}[h]
 \begin{centering}\vskip -0.5cm
 \includegraphics[width=1\linewidth]{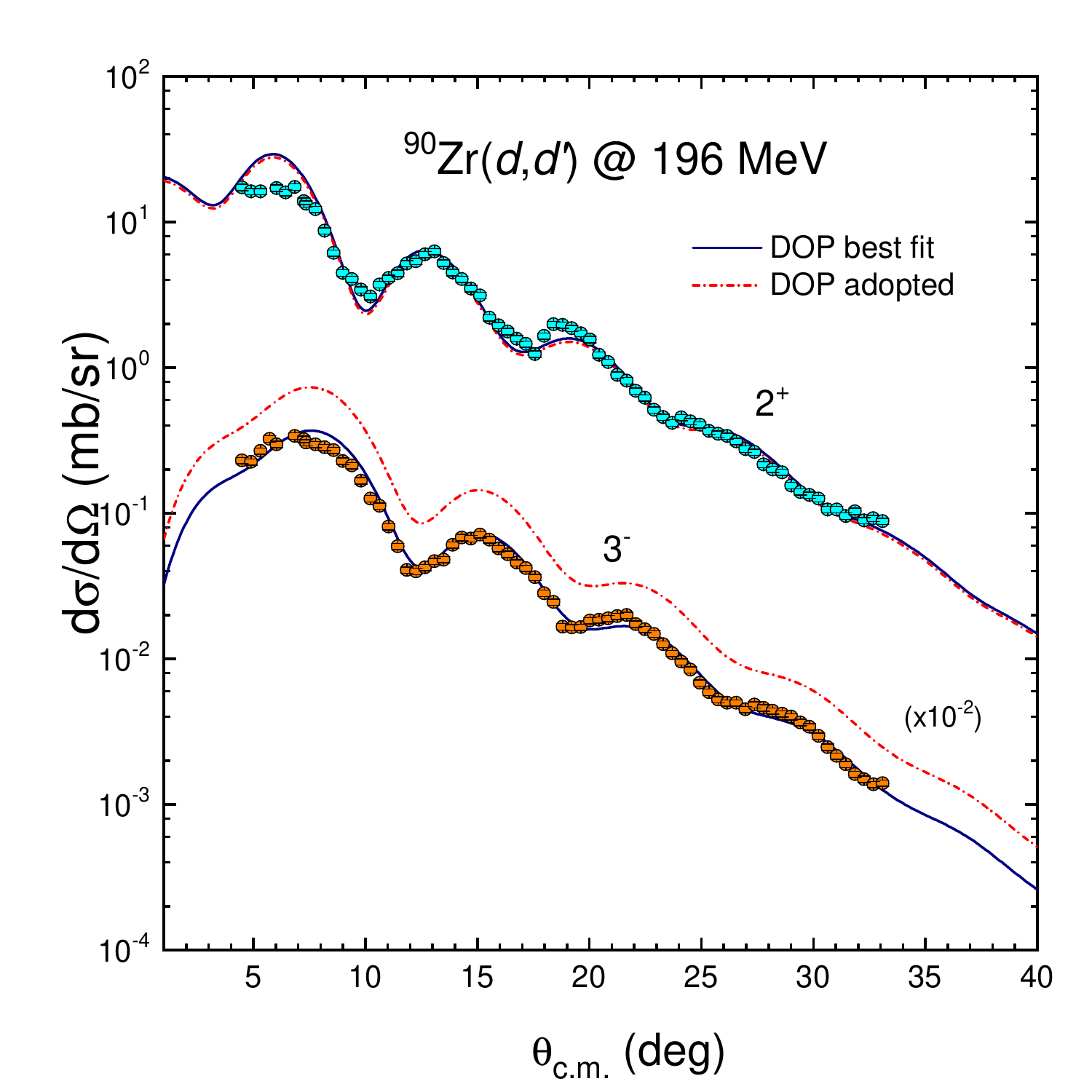}\vskip -0.2cm
 \caption{The same as Fig.~\ref{f4} but for inelastic $^{90}$Zr$(d,d')$ scattering data.} \label{f5}
 \end{centering}
 \end{figure}

The DWBA has been proven to be a reliable approximation for the direct reaction studies at energies 
around or above 100 MeV/nucleon. Therefore, a good agreement of the $(d,d')$ cross section calculated 
in the DWBA with $(d,d')$ data measured for 2$^+_1$ and 3$^-_1$ states of target must be a prerequisite 
for the validation of the phenomenological OP (\ref{eq1})-(\ref{eq6}). In the present work, the DWBA 
analysis of  inelastic deuteron scattering off  $^{90}$Zr and $^{116}$Sn targets was chosen as test ground 
for the phenomenological \dA OP. The DWBA results obtained for $^{116}$Sn and $^{90}$Zr targets 
are shown in Figs.~\ref{f4} and \ref{f5}, respectively. At variance with light targets, the DWBA cross sections 
given by the collective model inelastic FF based on the best-fit OP and adopted $B(E2)_{\rm adopted}$ 
values \cite{Raman01,nndc} agree well with the measured 2$^+_1$ angular distribution for both $^{90}$Zr and 
$^{116}$Sn targets. The best  DWBA fit yields $B(E2)_{\rm DOP}$ values around 5\% and 15\% larger 
than $B(E2)_{\rm adopted}$ values for $^{90}$Zr and $^{116}$Sn, respectively (see Table~\ref{t3}).

The situation is different, on the other hand, for the 3$^-_1$ angular distributions measured for these
targets. While the DWBA calculation using the collective model inelastic FF based on the adopted 
$B(E3)_{\rm adopted}$ value \cite{Kibedi02} describes well the $(d,d')$ data measured for the 3$^-_1$ 
state of  $^{116}$Sn, the same prescription overestimates the 3$^-_1$ angular distribution measured 
for the $^{90}$Zr target over the whole angular range. Given $B(E3)_{\rm adopted}\approx 0.098\ e^2$b$^3$ 
adopted from $(e,e')$ data, and the empirical $B(E3)$ values of 0.037 to 0.079 $e^2$b$^3$ deduced from 
inelastic nucleon- and light-ion scattering data  (see Table~V in Ref.~\cite{Kibedi02}), such a disagreement 
with the 3$^-_1$ angular distribution for $^{90}$Zr would not be unexpected. 
Indeed, the best DWBA fit  to the measured 3$^-_1$ cross sections yields a lower value 
of $B(E3)_{\rm DOP}\approx 0.051\ e^2$b$^3$. 

 \section{DWBA analysis based on the folded \dA potential}\label{sec3}
The microscopic description of the \AA interaction is usually based on a chosen effective pairwise 
nucleon-nucleon (NN) interaction $v$ between projectile nucleon and bound nucleon of target. 
The density dependence of $v$ presumably takes into account the three-nucleon interaction and 
higher-order NN correlations. Among such models, the double-folding model (DFM) has been  
used widely to calculate the $\alpha$-nucleus and heavy-ion OP \cite{Sat79,Bra97,Kho00}. 
The established success of the DFM in describing the observed elastic scattering for 
many \AA systems indicates that it produces the dominant part of the \AA OP. Within 
the DFM, the \dA scattering potential can be formally determined \cite{Kho00} as 
\begin{equation}
 U_F=\langle dA' |V_{d-A}|dA\rangle= \mathcal{A}\sum_{i\in d}\sum_{j\in A}
   \sum_{j'\in A'} \langle ij' |v_{\rm NN}|ij\rangle, \label{df1}
\end{equation}   
where the diagonal ($A'=A$ ) and nondiagonal ($A'\neq A$) matrix elements describe  
elastic and inelastic \dA scattering, respectively. The antisymmetrization $\mathcal{A}$ 
accounts for all single-nucleon exchanges between deuteron and target, giving rise to a 
nonlocal exchange term of the potential (\ref{df1})  
\begin{equation}
U_F=\sum_{i\in d}\sum_{j\in A}\sum_{j'\in A'} [\langle ij'|v_{\rm D}|ij\rangle 
 +\langle ij'|v_{\rm EX}|ji\rangle], \label{df2}
\end{equation}
where $v_{\rm D(EX)}$ is the direct (exchange) part of $v$. Given the nonlocal exchange potential, 
one has to solve an integro-differential OM equation involving a nonlocal kernel of the \dA OP,
which has not been done so far due the complexity of the nonlocal DFM computation. In fact, the exact 
solution of the OM equation with the nonlocal folded OP was obtained only for elastic nucleon scattering 
(see, e.g., Ref.~\cite{Kho24} and references therein). Like previous DFM calculations \cite{Kho97,Kho00},
we have used in the present work the well-proven local approximation for the exchange potential,
based on the WKB approximation for the change in the \dA relative motion wave function induced 
by the exchange of spatial coordinates of each interacting nucleon pair (see more details in 
Ref.~\cite{Kho07r}). In this case, both the direct and exchange terms of the \dA potential (\ref{df2}) 
are evaluated  \cite{Kho00} by folding the deuteron and target densities of with the chosen effective 
NN interaction $v$. The accuracy of such a WKB approximation was shown again in a recent OM study 
of elastic \nA scattering \cite{Kho24}. With a localized exchange potential, not only the OM calculation 
is much simpler, but also the comparison with the (local) phenomenological OP (\ref{eq1})-(\ref{eq6}) 
is more direct which is essential for the present study.    

 \subsection*{Effective density-dependent NN interaction and nuclear densities}
The  CDM3Y6 density-dependent version of the M3Y interaction \cite{Kho97} is used in 
the present DFM calculation of the \dA potential (\ref{df1})-(\ref{df2}). The real density dependence 
of the CDM3Y6 interaction was adjusted by a realistic HF description of nuclear matter, with nuclear 
incompressibility $K\approx 252$ MeV \cite{Kho97}. The imaginary density dependence of the CDM3Y6 
interaction was introduced in Ref.~\cite{Cuong10} to reproduce the Brueckner–Hartree–Fock results 
for nucleon OP in nuclear matter by Jeukenne, Lejeune and Mahaux (the JLM potential) \cite{JLM77}. 
This (complex) density-dependent CDM3Y6 interaction was successfully used in the DFM calculation 
of the $\alpha+^{208}$Pb OP and inelastic scattering FF \cite{Cuong10} for the multipole 
decomposition analysis of  $(\alpha,\alpha')$ data measured for isoscalar giant resonances 
of $^{208}$Pb at 97 MeV/nucleon \cite{Uchida04}. We note that prior to the present 
work, the only DFM calculation of the \dA OP was done 42 years ago by Cook \cite{Cook82} 
using the density independent M3Y interaction.  

Beside the effective NN interaction, the nuclear densities of projectile and target are essential
inputs for the DFM calculation. In the present work, we have chosen the deuteron density given by
the exact solution of the two-body problem using the Argonne V18 interaction as the free NN interaction 
\cite{Wir95}. For the DFM calculation of the diagonal ($A'=A$ ) elastic \dA potential, the  
Fermi distribution has been adopted for the ground-state (g.s.) densities of  target nuclei:  
\begin{equation}
 \rho_0(r)=\rho_0/[1+\exp((r-c)/a)], \label{df3}
\end{equation}
with the parameters $\rho_0,\ c$ and $a$ listed in Table~\ref{t4}. 
\begin{table}[h!]
		\begin{center}
		\caption{Parameters of the Fermi distribution (\ref{df3}). } \label{t4}\vskip 0.3 cm
		\begin{tabular}{|c|c|c|c|c|c|c|} \hline
\multicolumn{7}{|c|}{Nuclear density distribution}\\ \hline
Target & $^{24}$Mg &$^{28}$Si & $^{58}$Ni & $^{90}$Zr & $^{116}$Sn & $^{208}$Pb \\
\hline
$\rho_0$ (fm$^{-3}$) & 0.170 & 0.167 & 0.176 & 0.165 & 0.154 & 0.157 \\
$c$ (fm) & 2.995 & 3.160 & 4.080 & 4.900 & 5.490 & 6.670 \\
$a$ (fm) & 0.478 & 0.523 & 0.515 & 0.515 & 0.515 & 0.545 \\
Reference & \cite{Hanna1979}&  \cite{Hanna1979} &  \cite{Kho01}&\cite{Sat94}&
\cite{Sat94}& \cite{Sat94} \\ \hline
\multicolumn{7}{|c|}{Charge density distribution} \\ \hline
Target & $^{24}$Mg &$^{28}$Si & $^{58}$Ni & $^{90}$Zr & $^{116}$Sn & $^{208}$Pb \\
\hline
$\rho_0$(fm$^{-3}$) & 0.0785 & 0.0841 & 0.0826 & 0.0726 & 0.0688 & 0.0628 \\
$c$ (fm) & 3.045 & 3.154 & 4.177 & 4.908 & 5.417 & 6.647 \\
$a$ (fm) & 0.523 & 0.523 & 0.523 & 0.523 & 0.523 & 0.523 \\ 
Reference & \cite{frickle1995}&  \cite{frickle1995} &  \cite{frickle1995}&\cite{frickle1995}&
\cite{frickle1995}& \cite{frickle1995} \\ \hline
\end{tabular}
\end{center}
\end{table}

The main input for the DFM calculation of the nondiagonal ($A'\neq A$) \dA potential 
(the folded inelastic scattering FF) is the nuclear transition density of the target excitation.   
The DFM method (\ref{df1})-(\ref{df2}) is also used to calculate the Coulomb inelastic FF 
microscopically by folding the deuteron charge density and charge transition density of the target 
excitation with the Coulomb force acting between two protons \cite{Kho00}. 
For the 2$^\lambda$-pole excitations under study, we adopt the so-called Bohr-Mottelson 
prescription \cite{BohrMott} for the nuclear- and charge transition densities 
\begin {equation}
 \rho_\lambda(r)=-\delta_\lambda\frac{d\rho_0(r)}{dr}\  \mbox{and}\ 
 \rho^{(\lambda)}_{\rm charge}(r)=-\delta_\lambda\frac{d\rho^{(0)}_{\rm charge}(r)}{dr}, \label{df4} 
\end{equation}
where the g.s. charge density $\rho^{(0)}_{\rm charge}(r)$ is determined using the same Fermi
distribution (\ref{df3}) with parameters listed in Table~\ref{t4}. The deformation length 
$\delta_\lambda$ is determined from the transition rate $B(E\lambda\uparrow)$ of the target 
excitation using the following relation
\begin {equation}
B(E\lambda)=e^2|M_\lambda|^2,\ \mbox{where}\ M_\lambda=\int_0^\infty r^{\lambda+2}
\rho^{(\lambda)}_{\rm charge}(r)~dr. \label{df5}
\end{equation}
At variance with the DOP prescription (\ref{eq11}), the deformation length $\delta_\lambda$
in Eq.~(\ref{df5}) is the scaling factor of the nuclear transition density (\ref{df4}) used in the 
folding calculation (\ref{df2}) of inelastic scattering FF, which is constrained by the considered 
$B(E\lambda)$ transition rate. 

\subsection*{Folded deuteron OP and inelastic scattering FF}
The deuteron breakup has been shown to affect significantly the deuteron OP at the surface 
\cite{Rawitscher74,Austern87}, and a WS surface term $W_D(r)$ has been added to the imaginary 
folded OP because the DFM does not take into account the dynamic polarization of the OP by 
the breakup effect. A slight renormalization of the strength of both the real and imaginary folded 
OP is also allowed in the OM analysis of elastic $(d,d)$ data, and the total OP is determined 
in the hybrid manner as 
\begin{equation}
U(r) = U^{(0)}_F(r) + iW_D(r)+V_{\ell s}(r)(\bm{\ell}\cdot\bm{s})+V_C(r) , \label{df6}
\end{equation}
where $U^{(0)}_F(r)=N_V V^{(0)}_F(r)+iN_W W^{(0)}_F(r)$ is the diagonal folded \dA 
potential (\ref{df2}). The surface and spin-orbit terms of the OP (\ref{df6}) are determined 
in the same way as in Eqs.~(\ref{eq4})-(\ref{eq5}). The OM results obtained with the hybrid folded 
deuteron OP (\ref{df6}) are shown as solid lines in Figs.~\ref{f1} and \ref{f2}, with the best-fit 
$N_{V(W)}$ coefficients, parameters of the surface and spin-orbit terms of the OP listed 
in Table~\ref{t5}.
\begin{table}[h]
	\begin{center}
	\caption{Best-fit $N_{V(W)}$ coefficients, parameters of the surface and spin-orbit terms of the 
	hybrid folded deuteron OP (\ref{df6}). Because of spin convention, the $V_{\ell s}$ value must be 
	divided by 2 when used in the numerical input of the code ECIS97 \cite{ecis}. The errors were 
	deduced from the weight of each parameter in the covariant multi-parameter $\chi^2$-search, with 
	$W_D,\ a_D,\ V_{\ell s}$, and $a_{\ell s}$ kept fixed during the search.}
	\vskip 0.3 cm \label{t5}
	\begin{tabular}{|c|c|c|c|c|c|c|} \hline
Target & $^{208}$Pb &$^{116}$Sn & $^{90}$Zr & $^{58}$Ni & $^{28}$Si & $^{24}$Mg \\ \hline
$N_V$ & $0.98\pm 0.01$ & $0.94\pm 0.01$ & $0.98\pm 0.01$ & $0.98\pm 0.01$ 
 & $0.95\pm 0.01$ & $0.95\pm 0.01$ \\
$N_W$ & $1.01\pm 0.02$ & $1.04\pm 0.02$ & $1.00\pm 0.01$ & $1.03\pm 0.03$ 
 & $1.09\pm 0.03$ & $1.00\pm 0.03$ \\ \hline
$W_D$ (MeV) & 7.60 & 7.60   & 7.60 &7.60 & 7.60  & 7.60 \\
$r_D$ (fm) & $0.70\pm 0.09$ & $0.70\pm 0.04$  & $0.55\pm 0.03$ & $0.92\pm 0.04$ 
 & $0.54\pm 0.07$  & $0.75\pm 0.08$ \\
$a_D$ (fm) & 0.65 & 0.65  & 0.65  & 0.65  & 0.65  & 0.65\\ \hline
$V_{\ell s}$ (MeV) & 4.22 & 4.22 & 4.22 & 4.22 & 4.22 & 4.22 \\
$r_{\ell s}$ (fm) & $1.03\pm 0.01$ & $1.00\pm 0.02$  & $1.03\pm  0.02$ & $1.05\pm 0.01$
  & $1.03\pm 0.02$  & $0.82\pm 0.02$ \\
$a_{\ell s}$ (fm) & 0.85 &  0.85  & 0.85  & 0.85  & 0.85  & 0.85 \\ \hline
		\end{tabular}
	\end{center}
\end{table}
With the surface absorption taken into account by $W_D(r)$ term of the imaginary deuteron OP (\ref{df6}),  
the impact of the deuteron breakup on the complex folded \dA OP seems to be small, with the obtained
$N_{V(W)}$ coefficients being quite close to unity (see Table~\ref{t5}).     

\begin{figure}
 \begin{centering}\vskip -0.5cm
 \includegraphics[width=1\linewidth]{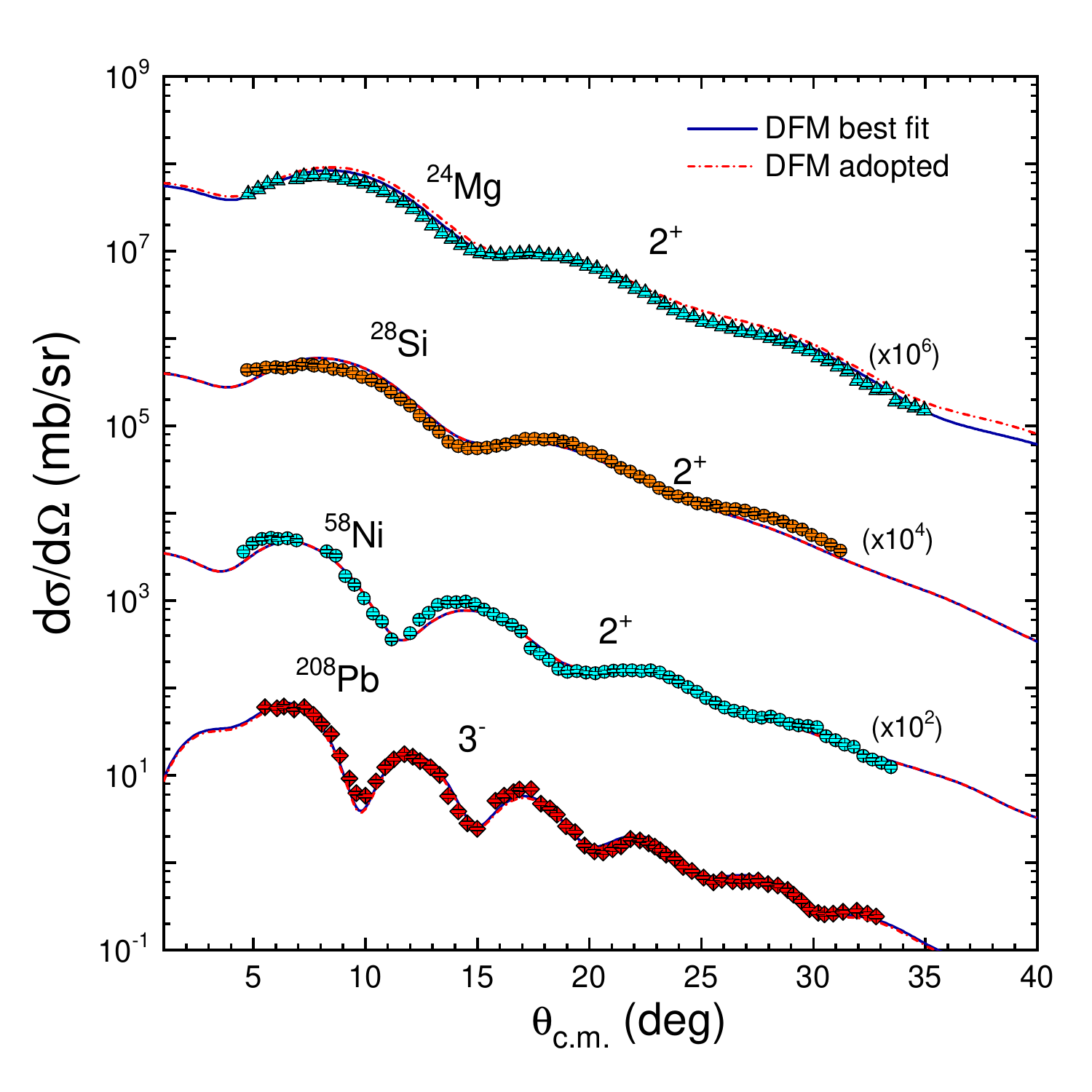}\vskip -0.2cm
 \caption{The same as in Fig.~\ref{f3} but obtained with the hybrid folded inelastic FF (\ref{df7}).
 The DWBA results based on the adopted $B(E\lambda)_{\rm adopted}$ and best-fit 
$B(E\lambda)_{\rm DFM}$ values (see Table~\ref{t3}) are shown as dashed and solid lines, 
respectively.} \label{f6}
 \end{centering}
 \end{figure}

The total inelastic $(d,d')$ scattering FF is also determined in the hybrid manner as  
\begin{equation}
F_\lambda(r)=U^{(\lambda)}_F(r)+F^{(\lambda)}_{W_D}(r) + 
F^{(\lambda)}_{\ell s}(r)(\bm{\ell}\cdot\bm{s})+F^{(\lambda)}_C(r),  \label{df7}
\end{equation}
where the folded nuclear $U^{(\lambda)}_F(r)=V^{(\lambda)}_F(r)+iW^{(\lambda)}_F(r)$ and Coulomb 
$F^{(\lambda)}_C(r)$ terms of the inelastic scattering FF are kept unchanged (as given by the DFM calculation) 
in the DWBA analysis of $(d,d')$ data. The surface and spin-orbit terms of the FF (\ref{df7}) are determined 
by the DOP method (\ref{eq9})-(\ref{eq10}), using the same deformation length $\delta_\lambda$ 
as that used for the nuclear- and charge transition densities (\ref{df4}). 
The DWBA results obtained with the hybrid folded inelastic FF (\ref{df7}) for $(d,d')$ scattering on 
$^{24}$Mg, $^{28}$Si, $^{58}$Ni, and $^{208}$Pb targets are shown in Fig.~\ref{f6}. Unlike the 
DWBA description of the 2$^+_1$ cross sections measured for light $^{24}$Mg and $^{28}$Si targets 
given by the collective model FF (\ref{eq9})-(\ref{eq12}) shown in Fig.~\ref{f3}, the hybrid folded 
FF (\ref{df7}) reproduces these same $(d,d')$ data very well using the $B(E2)_{\rm adopted}$ 
values. The oscillation pattern of the 2$^+_1$ cross section observed for these targets is also better 
reproduced by the hybrid folded FF compared to the collective model FF, and the best-fit $B(E2)_{\rm DFM}$ 
value agrees nicely with the $B(E2)_{\rm adopted}$ value as shown in Table~\ref{t3}.  

It should be noted here that the deformation parameter $\beta_2$ is obtained in one case from the OP geometry  
using the DOP method (\ref{eq9})-(\ref{eq10}), while in the other case it is obtained (\ref{df4}) from the geometry 
of the g.s. density. A more accurate comparison could be made based on the deformation length $\beta_2R$ obtained 
from both geometries or even better to compare the full integral of both the folded inelastic FF and collective 
model FF determined with the respective deformation parameters. 

\begin{figure}
 \begin{centering}\vskip -0.5cm
 \includegraphics[width=1\linewidth]{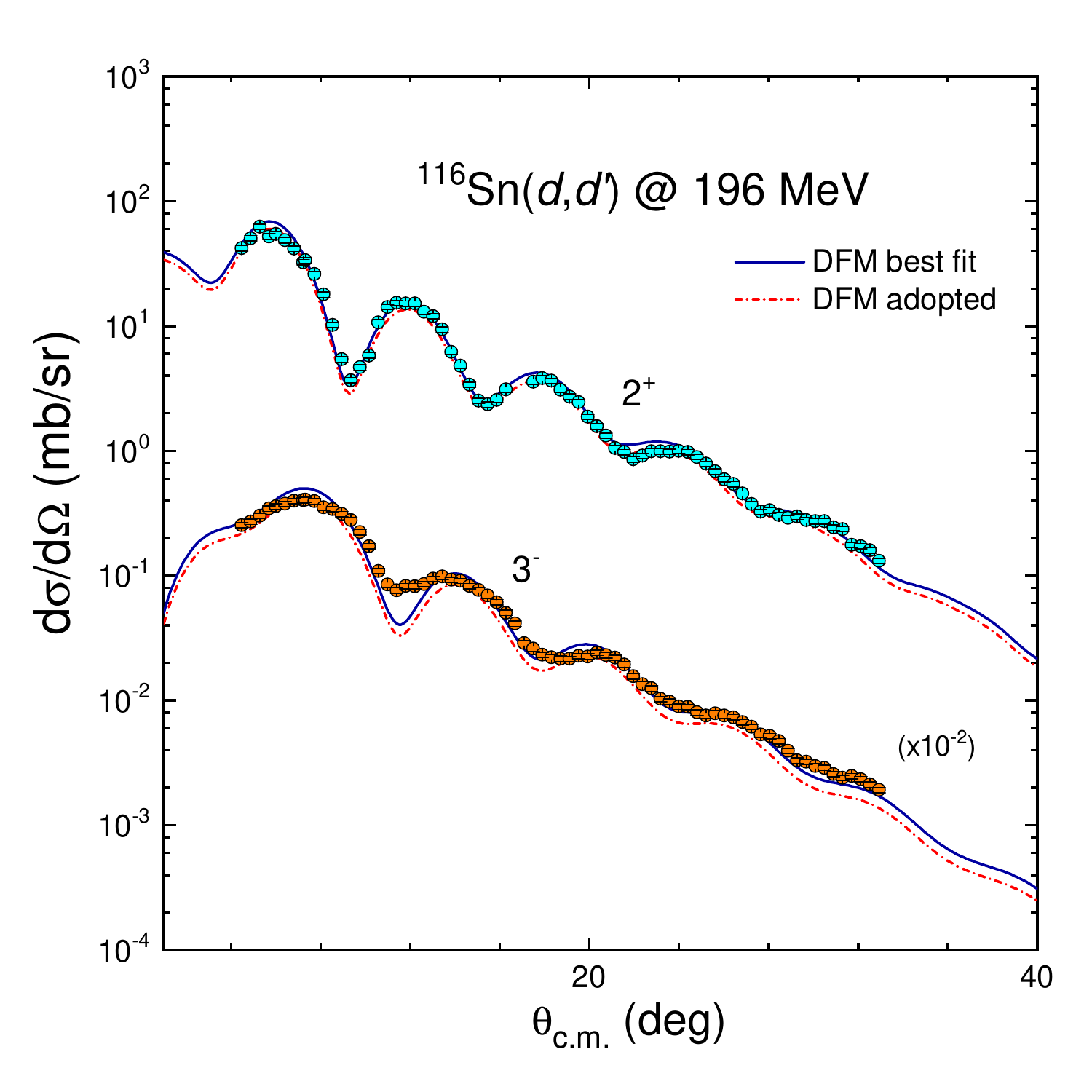}\vskip -0.2cm
 \caption{The same as in Fig.~\ref{f4}, but obtained with the  hybrid folded inelastic FF (\ref{df7}). 
The DWBA results based on the adopted $B(E\lambda)_{\rm adopted}$ and best-fit 
$B(E\lambda)_{\rm DFM}$ values (see Table~\ref{t3}) are shown as dashed and solid lines, 
 respectively.} \label{f7}
 \end{centering}
 \end{figure}

\begin{figure}
 \begin{centering}\vskip -0.5cm
 \includegraphics[width=1\linewidth]{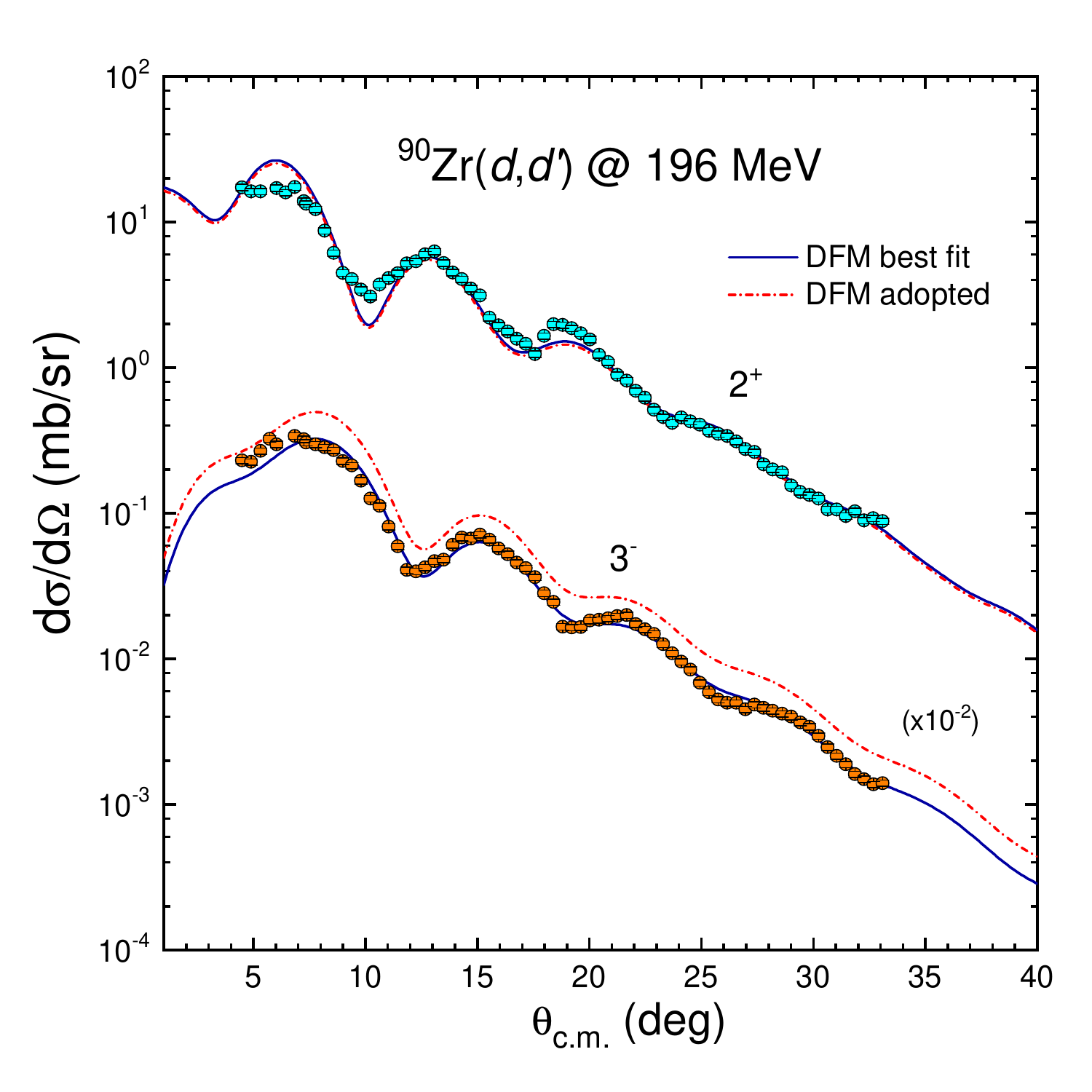}\vskip -0.2cm
 \caption{The same as in Fig.~\ref{f5}, but obtained with the hybrid folded inelastic FF (\ref{df7}). 
The DWBA results based on the $B(E\lambda)_{\rm adopted}$ and best-fit 
$B(E\lambda)_{\rm DFM}$ values (see Table~\ref{t3}) are shown as dashed and solid lines, 
respectively.} \label{f8}
 \end{centering}
 \end{figure}

The DWBA descriptions of the $2^+_1$ and $3^-_1$ angular distributions given by the hybrid folded 
inelastic FF (\ref{df7}) for $^{116}$Sn and $^{90}$Zr targets are shown in Figs.~\ref{f7} and \ref{f8}, 
respectively. One can see that the collective-model and hybrid folded inelastic scattering FF's give 
nearly equivalent DWBA descriptions of $(d,d')$ data measured for the $2^+_1$ excitation of these 
nuclei. The best-fit $B(E2)_{\rm DOP}$ and $B(E2)_{\rm DFM}$ values obtained for $^{90}$Zr 
and $^{116}$Sn targets are larger than the adopted values by around 5\% and 15\%, 
respectively (see Table~\ref{t3}).    

Like the DWBA results given by the collective-model FF shown in Figs.~\ref{f4} and \ref{f5},
the DWBA results given by the hybrid folded FF (\ref{df7}) based on the $B(E3)_{\rm adopted}$ 
value \cite{Kibedi02} reasonably reproduce the $(d,d')$ data measured 
for the 3$^-_1$  state of  $^{116}$Sn, but overestimate the data measured for the 3$^-_1$ state 
of $^{90}$Zr over the entire angular range. The best-fit $B(E3)_{\rm DFM}$ value is around 
35\% smaller than $B(E3)_{\rm adopted}$ value from $(e,e')$ data, but in a good agreement 
with the empirical $B(E3)$ values deduced from inelastic nucleon- and light-ion scattering data 
\cite{Kibedi02}. 

We note further that the best-fit $B(E3)_{\rm DOP}$ value obtained for the 3$^-_1$ state of $^{90}$Zr 
using the collective-model FF is about 20\% lower than the best-fit $B(E3)_{\rm DFM}$ value. 
A similar trend was also found for $3^-_1$ states of $^{116}$Sn and $^{208}$Pb (see Table~\ref{t3}) 
which shows that the DOP method consistently gives a lower $B(E3)$ value compared to that given 
by the DFM approach, as discussed previously in Ref.~\cite{Horen93}. However, the DOP method seems 
to better reproduce $B(E3)_{\rm adopted}$ value from the Coulomb excitation data 
\cite{Kibedi02}.

\section*{Summary}
Elastic and inelastic deuteron scattering have been measured off $^{24}$Mg, $^{28}$Si, $^{58}$Ni, $^{90}$Zr, 
$^{116}$Sn, and $^{208}$Pb at an energy of 98 MeV/nucleon.  The measured $(d,d)$ and $(d,d')$ 
angular distributions were analyzed within the OM and DWBA using the phenomenological deuteron OP 
associated with the collective model of nuclear scattering, and the hybrid potential model for the deuteron 
OP and inelastic FF based on the microscopic DFM calculation.  The $E2$ and $E3$ transition rates 
of the $2^+_1$ and $3^-_1$ excitations of these target nuclei have been deduced from the best DWBA 
fits to the $(d,d')$  scattering data, which agree reasonably with the adopted $B(E\lambda)$ 
values \cite{Raman01,Kibedi02}.

The results of the OM analysis of elastic $(d,d)$ data using the hybrid folded OP show that the deuteron 
breakup does not affect significantly the volume part of the deuteron OP given by the DFM calculation. 
However, the breakup effect seems to imply an enhanced absorption at the surface, which can be taken 
into account effectively by a surface WS potential added to the imaginary deuteron OP.

While both potential models describe equally well the $(d,d)$ and $(d,d')$ angular distributions
measured for medium- and heavy-mass target 
nuclei, the DWBA calculation using the collective-model inelastic FF (DOP) gives a poorer description 
of the $2^+_1$ angular distribution measured for the light-mass $^{24}$Mg and $^{28}$Si nuclei, in comparison 
with the DWBA description of the same data using the semi-microscopic folded inelastic FF (DFM).  

The DWBA analysis of the $3^-_1$ angular distribution measured for $^{90}$Zr target indicates 
that $B(E3)_{\rm adopted}$ value from $(e,e')$ data \cite{Kibedi02} is too high, and the 
DWBA results given by the two potential models agree well with $(d,d')$ scattering data, with the 
best-fit $B(E3)$ values close to those deduced from inelastic nucleon- and light-ion scattering data. 

A hindrance of the $E3$ transition rate determined by the collective-model inelastic FF compared 
to that determined by the inelastic folded FF was also found, which illustrates the inconsistency 
between the DWBA description of nuclear excitation with $\lambda\gtrsim 3$ based on the DOP 
approach and that based on the DFM approach discussed earlier by 
Beene \emph{et al.} \cite{Horen93}.

\section*{Acknowledgment}
We acknowledge the efforts of the RCNP staff in providing high-quality deuteron beams required for 
these measurements, and thank A. Okamoto (Konan), T. Sako (Kyoto) and K. Schlax (Notre Dame) 
for their assistance with the experiment.  This work has been supported in part by the National Science 
Foundation (Grants No. PHY-1068192, No. PHY-2011890, and No. PHY-2310059). Two of us 
(D.C.C. and D.T.K.) were supported by the National Foundation for Science and Technology 
Development of Vietnam (NAFOSTED Project No. 103.04-2021.74).

\end{document}